\documentclass[12pt,a4paper]{article}
\usepackage{graphicx}
\usepackage{color}

\usepackage{bm,enumerate,amsmath,amssymb,amsthm}
\usepackage{epsfig}
\usepackage{cite}
\usepackage{algorithm}
\usepackage{algpseudocode}
\usepackage{txfonts}
\usepackage{subfigmat}
\usepackage{figsize}
\usepackage{here}
\usepackage{listings}
\usepackage{braket}
\usepackage{comment}
\usepackage{lscape}
\usepackage{bigints}

\usepackage{arxiv}
\usepackage[utf8]{inputenc} 

\theoremstyle{definition}

\newtheorem*{theorem*}{Theorem}

\newtheorem*{definition*}{Definition}

\renewcommand{\headeright}{}

\title{A regularization method for quantum neural networks using data symmetry}

\author{
  Hiroshi Ohno\\
  Toyota Central R\&D Labs., Inc.\\
  Aichi, Japan\\
  \texttt{oono-h@mosk.tytlabs.co.jp}\\
}

\date{\empty}

\begin{document}

\maketitle

\begin{abstract}
  Leveraging data symmetries has recently become a key strategy in quantum neural networks (QNNs) to improve training efficiency.
  In this study, we propose a symmetry-informed regularization method for QNNs based on an input density matrix.
  By introducing a penalty term that encourages the model to align with data symmetry, our method enables improved training speed.
  This symmetry-based regularization is simple to implement and does not require an explicitly specified symmetry group, although it requires access to training samples or to the input distribution from which an empirical or theoretical input density matrix can be constructed.
  We evaluate the method through numerical experiments on both classification tasks and quantum generative adversarial networks.
  In the small-scale classification experiments, the regularizer produced modest improvements in early-stage convergence and test loss.
  Our findings highlight the potential of symmetry-aware regularization in enhancing the performance of QML models.
  \keywords{Symmetry \and Quantum neural networks \and Penalty term}
\end{abstract}

\section{Introduction}
Quantum machine learning (QML) has emerged as one of the most promising applications of quantum computing, leveraging recent advances in quantum hardware, particularly noisy intermediate-scale quantum (NISQ) devices \cite{bharti2022} and variational quantum algorithms \cite{cerezo2021}.
QML has been applied to a variety of machine learning tasks \cite{havlicek2019,huang2021,owen2023,yu2024} such as generative modeling, classification, and regression, mirroring capabilities in classical machine learning.
In practical settings, QML models are typically trained using a quantum-classical hybrid approach \cite{peruzzo2014}, where parameterized quantum circuits are optimized using classical algorithms.
However, due to the need for repeated quantum measurements and the overhead of encoding classical data into quantum states, training speeds tend to be low, especially for large-scale QML models.
In this study, we aim to address two key challenges in QML: improving training efficiency.
To this end, we focus on exploiting symmetries in the input data as a strategy for enhancing training efficiency.

Symmetries play a fundamental role in physical systems, particularly in connection with conserved quantities.
In the context of quantum chemistry, several studies have investigated how symmetries can be exploited within variational algorithms \cite{picozzi2023,dibyendu2025}.
In QML, data symmetries are utilized to enhance both the trainability and generalization capabilities of models \cite{larocca2022,meyer2023,schatzki2024,nguyen2024}\footnote{
  In cases where there is no inherent data symmetry, for example, in the Grover's search algorithm \cite{grover1996}, the algorithm introduces symmetry into the computation, resulting in a quadratic speedup.
}.
As a form of inductive bias, Ref. \cite{larocca2022} proposed a method for embedding intrinsic invariances of the training data into quantum neural networks (QNNs).
Ref. \cite{schatzki2024} demonstrated that permutation-equivariant QNNs can avoid barren plateaus, rapidly reach an overparameterized regime, and generalize well from small training datasets.
At the same time, symmetry constraints reduce the dimension of the accessible hypothesis class.
If the imposed symmetry is incompatible with the learning task, such a restriction may lead to insufficient expressibility.
Ref. \cite{nguyen2024} further explored the integration of symmetries into QNNs by designing network architectures aligned with the symmetry group of the training data.
In particular, the authors constructed equivariant quantum convolution and pooling layers that respect the underlying symmetries of the data.
These three studies focused on label symmetries expressed in the form $ f(U \rho U^{\dagger}) = f(\rho) $, where $ U $ is a unitary operator acting on the quantum state $ \rho $, and $ f $ denotes the model function.
This formulation ensures invariance of the model output under symmetry transformations.
In Ref. \cite{meyer2023}, a different approach was taken based on representation theory.
The authors proposed a method called gate symmetrization, in which a standard gate set is transformed into an equivalent set that respects the symmetry of the task.
Experimental results on toy problems showed improved generalization performance, demonstrating the effectiveness of symmetry-aware ansatz design.

Based on the aforementioned previous studies, we summarize the effects of data symmetries as follows:
In models with inductive biases (i.e., data symmetries), the parameter search space is constrained (limited) by these symmetries, which accelerates the training process.
Since redundant data generated through symmetry transformations can be disregarded, the model can be trained effectively with relatively few data samples, leading to improved generalization performance.
Furthermore, vanishing gradients in the cost function during training are alleviated by removing redundant directions (e.g., zero-gradient directions) introduced by symmetries, thereby helping to avoid barren plateaus.
However, excessive symmetries (i.e., overly strong inductive biases) can overly restrict the search space, resulting in poor generalization and limited expressibility.
Therefore, to prevent excessive symmetries, it is necessary to introduce mechanisms for symmetry breaking, such as gates.

As an alternative approach to introducing inductive biases, symmetry can be incorporated into the training cost function in the form of a penalty term (regularization) \cite{kuroiwa2021}.
In Ref. \cite{kuroiwa2021}, for variational quantum eigensolvers, a penalty term reflecting the symmetry of the wave function was introduced.
This penalty term allows the training algorithm to avoid exploring states outside the desired subspace.
A balancing parameter (i.e., a hyperparameter) controls the strength of the penalty term, making it easier to avoid the imposition of excessive symmetries.

In this study, we adopt state symmetry ($ U \rho U^{\dagger} = \rho $) and a penalty term, and propose a heuristic regularization method based on input data symmetry for QNNs.
The objective of this regularization is to improve both training efficiency and test loss (i.e., generalization performance).
In the proposed method, to align the symmetries of the model and the data, we introduce a penalty term based on the distance between the model (a parameterized unitary) and a unitary matrix derived from the input data symmetry.
The overall cost function is composed of the original cost and the penalty term.
Importantly, prior knowledge of the underlying data symmetry group is not required\footnote{
  In this study, we use the term ``prior knowledge'' to refer to properties of (classical) data that are known before any data preprocessing is performed.
  When data symmetry is known as prior knowledge, methods that exploit it can typically outperform the proposed method.
}.
To evaluate the effectiveness of the proposed regularization, we conduct two types of experiments: classification and generative tasks.
For the generative tasks, we employ quantum generative adversarial networks (QGANs) \cite{huang2021} to generate handwritten digits.
The experimental results provide empirical evidence supporting the effectiveness of the proposed regularization method.

Our main contributions are summarized as follows:
\begin{itemize}
\item We propose a simple regularization method for QNNs based on input data symmetry.
\item Numerical experiments on classification and generative tasks demonstrate improvements in training speed.
\end{itemize}

The remainder of this paper is organized as follows.
Section \ref{sec2} describes the proposed method.
Section \ref{sec3} details the numerical experiments conducted on classification and generative tasks, along with the corresponding results for training and test losses.
For the generative task, we show the images generated by models with and without data symmetry.
Finally, Section \ref{sec4} presents our conclusions and discusses potential directions for future work.

\section{Methods}\label{sec2}
In this section, we describe a regularization method for QML as follows.
\begin{enumerate}
\item Training models (quantum circuits)
\item Data symmetry for regularization
\item A cost function with regularization for training
\end{enumerate}

\subsection{Training model}
A training model (quantum circuit) $ U $ for training is constructed as follows:
\begin{equation}\label{eq1}
  \begin{split}
    U(x, \theta) &= U_{enta} U(\theta_{d}) U_{enta} U(\theta_{d-1}) \cdots U_{enta} U(\theta_{1}) U_{encoding}(x)\\
    &= \left( \prod_{l=1}^{d} U_{enta} U(\theta_{d - l + 1}) \right) U_{encoding}(x),
  \end{split}
\end{equation}
where $ x $ denotes an input data, $ \theta $ denotes a set of parameter vectors for training, e.g., $ \theta = \{ \theta_{1}, \ldots, \theta_{d} \}$, $ U_{enta} $ denotes an entangling circuit which is realized by circuits composed of two-qubit gates (such as CZ and CNOT gates) based on a configuration \cite{sim2019}, $ U(\theta_{l}) $ denotes the $l$-th parameterized circuit (parameterized unitary matrix), $ U_{encoding} $ denotes a data-encoding circuit, and $ d $ is a circuit depth.
Here, $ W(\theta) \coloneqq \prod_{l=1}^{d} U_{enta} U(\theta_{d - l + 1}) $.

In this study, $ U(\theta_{l}) $ and $ U_{encoding} $ are constructed as follows:
\begin{equation}\label{eq2}
  U(\theta_{l}) = RY(\theta_{l,1}) \,RZ(\theta_{l,n+1}) \otimes RY(\theta_{l,2}) \,RZ(\theta_{l,n+2}) \cdots \otimes RY(\theta_{l,n}) \,RZ(\theta_{l,2n}),
\end{equation}
and
\begin{equation}\label{eq3}
  U_{encoding}(x) = RY(x_{1}) \otimes RY(x_{2}) \cdots \otimes RY(x_{n}).
\end{equation}
The total number of trainable parameters is $ 2 n d $.
With an appropriate entangling topology and sufficiently large circuit depth, this gate set is universal for $ n $-qubit unitary operations.

For an $n$-qubit system, the output state $ \ket{\psi} $ is defined as
\begin{equation}\label{eq4}
  \ket{\psi(x, \theta)} \coloneqq U(x, \theta) \ket{0}^{\otimes n}.
\end{equation}
The model output $ y(x) $ is then calculated by
\begin{equation}\label{eq5}
  y(x) = \braket{\psi(x, \theta) | \, \mathcal{O} \, | \psi(x, \theta)},
\end{equation}
where $ \mathcal{O} $ denotes an observable.

\subsection{Data symmetry}
First, we describe the input density matrix $ \rho $ of input data $ x $ as defined by
\begin{equation}\label{eq6}
  \rho_{in} \coloneqq \mathbb{E}_{x} \left[ U_{encoding}(x) \ket{0}^{\otimes n} \bra{0}^{\otimes n} U_{encoding}^{\dagger}(x) \right],
\end{equation}
where the expectation $ \mathbb{E} $ is taken over input data $ x $.

For classification tasks, given an input data set with size $ M $, the density matrix is formulated by
\begin{equation}\label{eq7}
  \rho_{in} = \frac{1}{M} \sum_{j=1}^{M} U_{encoding}(x_{j}) \ket{0}^{\otimes n} \bra{0}^{\otimes n} U_{encoding}^{\dagger}(x_{j}).
\end{equation}
The computational complexity is $ \mathcal{O}(2^{2n}) $.
A practically scalable variant will likely require replacing the exact global symmetry extracted from the full input density matrix with an approximate structured symmetry obtained from local marginals, factorized approximations, or a restricted family of symmetry generators.

For generative models such as a generator in QGANs \cite{huang2021}, the density matrix is formulated by
\begin{equation}\label{eq8}
  \rho_{in} = \int_{x} U_{encoding}(x) \ket{0}^{\otimes n} \bra{0}^{\otimes n} U_{encoding}^{\dagger}(x) \, dP_{x},
\end{equation}
where $ dP_{x} $ denotes a measure of $ x $, providing noises as inputs of a generator.
This can also be applied when the input data distribution is known in classification tasks.

Additionally, when $ x $ is encoded to each qubit ($ x_{j} $, $ 1 \leq j \leq n $) and $ x_{j} $ are independent of each other, $ \rho_{in} $ is given by
\begin{equation}\label{eq9}
  \rho_{in} = \int_{x_{1}} U_{encoding}(x_{1}) \ket{0} \bra{0} U_{encoding}^{\dagger}(x_{1}) \, dP_{x_{1}} \otimes \, \cdots \, \otimes \int_{x_{n}} U_{encoding}(x_{n}) \ket{0} \bra{0} U_{encoding}^{\dagger}(x_{n}) \, dP_{x_{n}},
\end{equation}
where $ U_{encoding} $ is a single-qubit gate.

Next, we focus on the data symmetry such that $ \rho_{in} $ commutes with $ S $, i.e., $ [\rho_{in}, S] = 0 $, which is mathematically equivalent to $ S \rho_{in} S^{\dagger} = \rho_{in} $,\footnote{
  The proof is as follows:
  Assume that $ X $ and $ Y $ are a unitary matrix.
  \underline{ $ [X, Y] = 0 \, \rightarrow \, Y X Y^{\dagger} = X $ }:
  From $ [X, Y] = XY - YX = 0 $, $ YX = XY $.
  Then, $ Y X Y^{\dagger} = X Y Y^{\dagger} = X {\rm I} = X $, where $ {\rm I} $ is an identity matrix.\\
  \underline{ $ [X, Y] = 0 \, \leftarrow \, Y X Y^{\dagger} = X $ }:
  From $ Y X Y^{\dagger} = X $, $ Y X Y^{\dagger} Y = X Y $, $ Y X {\rm I} = X Y$, $ Y X = X Y $, and thus $ [X, Y] = 0 $.
} where $ S $ is a unitary matrix.
We can construct $ S $ using the spectral decomposition of $ \rho_{in} $ (an algorithm is shown in Appendix \ref{ap1}).

\subsubsection{Model with data symmetry}
For a model $ W $, we can also construct $ V $ such that $ [W, V] = 0 $, where $ V $ is a unitary matrix.
Here, $ V $ is not always equal to $ S $, and thus $ W $ does not possess the data symmetry.
In general, this can lead to slow training and poor generalization performance.
By incorporating the data symmetry (i.e., providing an inductive bias) into the model, it appears that we can address these issues.

First, for the model output $ y $, we convert the expectation representation to the trace ($ {\rm Tr} $) representation.
From Equation \ref{eq5}, $ y = f(x) = \braket{\psi(x, \theta) | \, \mathcal{O} \, | \psi(x, \theta)} $.
Here, $ \ket{x} \coloneqq U_{encoding}(x) \ket{0}^{\otimes n} $.
Then, $ f(x) = \braket{x | \, W^{\dagger} \mathcal{O} W \, | x} $.
We modify $ f(x) $ as follows:
\begin{equation}\label{eq12}
  \begin{split}
    f(x) &= \braket{x | \, W^{\dagger} \mathcal{O} W \, | x}\\
    &= \braket{x | \sum_{k} \ket{e_{k}} \bra{e_{k}} \, W^{\dagger} \mathcal{O} W \, | x}\\
    &= \sum_{k} \braket{x | e_{k}} \braket{e_{k} | \, W^{\dagger} \mathcal{O} W \, | x}\\
    &= \sum_{k} \braket{e_{k} | \, W^{\dagger} \mathcal{O} W \, | x} \braket{x | e_{k}}\\
    &= {\rm Tr}[ W^{\dagger} \mathcal{O} W \, \ket{x} \bra{x} ]\\
    &= {\rm Tr}[ W^{\dagger} \mathcal{O} W \, \rho(x) ],
  \end{split}
\end{equation}
where $ e_{k} $ denotes any orthonormal basis and $ \rho(x) \coloneqq \ket{x} \bra{x} $.

Here, we assume that the symmetry of model is different from that of data ($ V \neq S $), $ S \rho(x) S^{\dagger} = \rho(x) $ ($[ \rho(x), S ] = 0$), and $ V W V^{\dagger} = W $ ($[ W, V ] = 0 $).
Then, we use the cyclic property of trace for Equation \ref{eq12} as follows.
\begin{equation}\label{eq13}
  \begin{split}
    f(x) &= {\rm Tr}[ W^{\dagger} \mathcal{O} W \, \rho(x) ]\\
    &= {\rm Tr}[ (V W V^{\dagger})^{\dagger} \mathcal{O} (V W V^{\dagger}) \, S \rho(x) S^{\dagger} ]\\
    &= {\rm Tr}[ V W^{\dagger} V^{\dagger} \mathcal{O} V W V^{\dagger} \, S \rho(x) S^{\dagger} ]\\    
    &= {\rm Tr}[ \mathcal{O} V W V^{\dagger} S \rho(x) S^{\dagger} V W^{\dagger} V^{\dagger} ]\\
    &= {\rm Tr}[ \mathcal{O} T \rho(x) T^{\dagger} ],
  \end{split}
\end{equation}
where $ T \coloneqq V W V^{\dagger} S $.

When the symmetry of model is the same as that of data ($ V = S $) for Equation \ref{eq13}, since $ V = S $ and $ [W, V] = 0 $,
\begin{equation}\label{eq14}
  \begin{split}
    f(x) &= {\rm Tr}[ \mathcal{O} T \rho(x) T^{\dagger} ]\\
    &= {\rm Tr}[ \mathcal{O} V W \rho(x) W^{\dagger} V^{\dagger} ]\\
    &= {\rm Tr}[ \mathcal{O} W S \rho(x) S^{\dagger} W^{\dagger} ]\\
    &= {\rm Tr}[ \mathcal{O} W \rho(x) W^{\dagger} ]\\
    &= {\rm Tr}[ W^{\dagger} \mathcal{O} W \rho(x) ].
  \end{split}
\end{equation}
Thus, we obtain Equation \ref{eq12} again.
The penalty restricts the optimization toward the commutant of the selected symmetry operator $ S $.
This reduces the effective search space, although such a restriction is beneficial only when the selected symmetry is compatible with the target task.

\subsection{Symmetry-based penalty term}
For $ W(\theta) $, a penalty loss function $ C_{symm} $ regarding data symmetry is defined as
\begin{equation}\label{eq19}
  C_{symm} \coloneqq \| S W(\theta) S^{\dagger} - W(\theta) \|_{{\rm F}},
\end{equation}
where $ \| \cdot \|_{{\rm F}} $ denotes the Frobenius norm (A mismatch bound is shown in Appendix \ref{ap2}).
This penalty term multiplied by a positive constant $ C $ is added to the original cost function we optimize, which regularizes $ W(\theta) $.
Here, $ C $ is a balancing parameter, controlling the amount of data symmetry.
The stronger the symmetry is imposed on the unitary $ W(\theta) $, the narrower its parameter space becomes, and thus the training speed may be faster.

In the experiments, $ W(\theta) $ and $ C_{symm} $ were evaluated using full-matrix classical simulation.
A parameter shift-rule algorithm \cite{schuld2019,gacon2021} was exploited to optimize the cost function.

\section{Numerical experiments and results}\label{sec3}
To examine the effect of penalty term regarding data symmetry, we conducted two learning tasks: classification and generative tasks\footnote{
  The simulation experiments were performed under the following conditions.
  Quantum circuits were implemented by PennyLane \cite{bergholm2018}.
  Discriminator in QGAN was implemented by PyTorch \cite{paszke2019}.
  For quantum circuit simulations, noiseless and state-vector simulations were performed.
  For classification tasks, a stochastic gradient descent (SGD) optimizer based on the parameter-shift rule algorithm was used, with a learning rate of 0.01 and a momentum term of 0.5.
  For generative tasks, a SGD optimizer was used, with a learning rate of 0.01 for discriminators and 0.3 for generators of QGAN.
}.
For the generative tasks, we used QGANs \cite{huang2021}, where a generator ($ G $) in QGAN is realized by a quantum circuit, while a discriminator ($ D $) is a classical neural network with two hidden layers.

\subsection{Classification tasks}
For a classification task \cite{farhi2018,haug2024}, in the following, we prepared a training dataset $ \{x_{l}, y_{l}\}_{l=1}^{M} $, where $ x $ denotes a feature vector and $ y $ denotes a binary label $ y \in \{-1, 1\} $.

The datasets were created in the following way.
The Hamiltonian $ H $, serving as the generator of the target unitary matrix, was defined as follows \cite{haug2024}:
$ H = \sum_{j = 1}^{n - 1} X_{j} X_{j+1} + \sum_{j = 1}^{n - 1} Y_{j} Y_{j+1} + \sum_{j}^{n} h_{j} Z_{j} $, where $ X $, $ Y $, and $ Z $ are the Pauli matrices and $ h $ is a tunable parameter.
Using the angle encoding method \cite{schuld2018}, the input data $ x $ were encoded into quantum states using RY gates\footnote{
  The angle encoding initially produces a product state with zero entanglement entropy, allowing us to separate the entanglement generated by the variational circuit from that introduced during encoding.
}.
The output state then was obtained by $ \ket{\phi} = \left( \prod_{l=1}^{L} \prod_{k=1}^{K} \exp(-i \theta_{l,k} H) \right) \, RY(x)^{\otimes n} \, \ket{0}^{\otimes n} $.
And the measurement was performed by $ \hat{y} = \braket{\phi | \, Z(0) \, | \phi} $.
Finally, for the corresponding label $ \hat{y} $ to each $ x $, we defined the label $ y $ such that if $ \hat{y} \geq 0 $, $ y = 1 $; otherwise $ y = -1 $.

In the experiments, the input data $ x $ were sampled from a uniform distribution in the range $ [0, \, 2\pi) $; $ \theta $ was sampled from a uniform distribution in the range $ [0, \, 2\pi) $; $ h $ was randomly chosen from $ \{-1, \, 1 \} $; $ n = 5 $, $ L = 5 $, and $ K = 2 $.
The number of training data ($ M $) was 100, and the number of test data was 200.
The training dataset contained 53 positive samples (53\%) and 47 negative samples (47\%), whereas the test dataset contained 96 positive samples (48\%) and 104 negative samples (52\%).
According to Equation \ref{eq9}, since $ U_{encoding} $ is a RY gate and $ x $ is sampled independently from a uniform distribution with range $ [0, \, 2\pi) $, we can calculate as follows.
\begin{equation}\label{eq11}
  \rho_{in} = \frac{1}{2^{n}} {\rm I}_{2^{n}},
\end{equation}
where $ {\rm I}_{2^{n}} $ denotes an identity gate of size $ 2^{n} $.

\begin{figure}[htbp]
  \centering
  \begin{tabular}{c}
    \includegraphics[width=4cm]{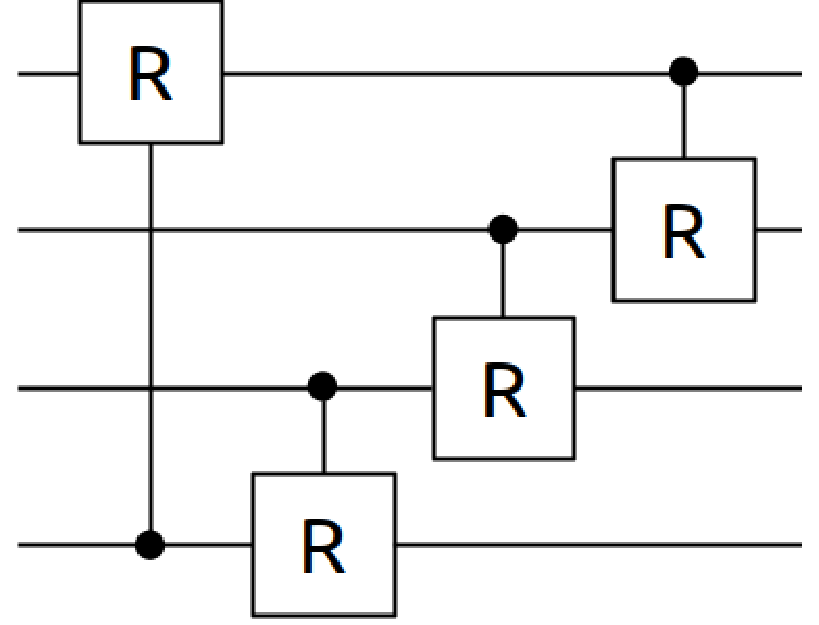}
  \end{tabular}
  \caption{Circuit-block configurations for four qubit entangling circuits. $ {\rm R} $ denotes a two-qubit gate implementing a controlled Pauli operator (X or Z).}\label{fig0}
\end{figure}

In the model (Equation \ref{eq1}), we adopted the circuit-block (CB) configuration and controlled Z gates (CZ) to construct $ U_{enta} $ (Figure \ref{fig0}) \cite{sim2019}.

The cost function $ C_{f} $ is defined as follows \cite{farhi2018}:
\begin{equation}
  C_{f} \coloneqq 1 - \frac{1}{M} \sum_{l = 1}^{M} y_{l} \, \hat{y}(x_{l}), 
\end{equation}
where $ \hat{y}(x_{l}) = \braket{\psi(x_{l}, \theta) | \, Z(0) \, | \psi(x_{l}, \theta)} $ (Equation \ref{eq5}).
Unlike classification accuracy, this continuous cost function accounts for both the correctness of the predicted sign and the magnitude of the model output.
Therefore, it was used as the optimization objective, while classification accuracy is additionally reported as an evaluation metric.

\subsection{Generative tasks}
\begin{figure}[htbp]
  \centering
  \begin{tabular}{c}
    \includegraphics[width=10cm]{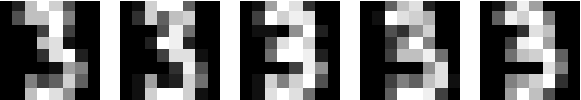}
  \end{tabular}
  \caption{Training image samples of handwritten digit '3' (five examples)}\label{fig3-1}
\end{figure}

For the QGAN construction, the number of qubits was five and two ancilla qubits (which determine the degree of non-linear mappings in the generators).
The generator circuit consists of RY and CZ gates, and for a given depth $ d $, it includes $ d $ repetitions of CZ and RY gates \cite{huang2021}.
Input noise sampled from a uniform distribution over the range $ [0, \, \pi/2) $ is fed into the RY gates of the generator circuit.
The training image data used the handwritten digits data set for optical recognition\footnote{
  https://archive.ics.uci.edu/dataset/80/optical+recognition+of+handwritten+digits .
}.
The image size was 64 bits as shown in Figure \ref{fig3-1}.
This experiment is a single-class generation task; only samples labeled as the digit ``3'' were used for training.
Following the patch-based QGAN architecture in Ref. \cite{huang2021}, we used eight quantum sub-generators. 
Each sub-generator consisted of five data qubits and two ancilla qubits and generated one 8-dimensional patch.
The eight patches were concatenated and reshaped into an $ 8 \times 8 $ image.

Training of QGAN was performed up to 500 iterations.
According to Equation \ref{eq9}, since a uniform distribution with range $ [0, \, \pi/2) $ is used, we obtain
\begin{equation}\label{eq10}
  \rho_{in} = \left( \begin{array}{cc}
    \frac{1}{2} + \frac{1}{\pi} & \frac{1}{\pi}\\
    \frac{1}{\pi} & \frac{1}{2} - \frac{1}{\pi}
  \end{array} \right)^{\otimes n}.
\end{equation}

For a generator cost function, a negative cost function $ -\mathbb{E}_{z}[ \log(D(G(z))) ] $ was used\footnote{
  In GAN, a generator cost function is defined as $ \log(1 - D(G(z))) $ \cite{goodfellow2014}. 
}, where $ z $ is an input noise

\subsection{Classification task results}
\begin{figure}[htbp]
  \centering
  \begin{minipage}{12cm}
    \centering
    \SetFigLayout{2}{2}
    \subfigure[d = 2 (e), up to 200 epochs]{\includegraphics[width=5cm]{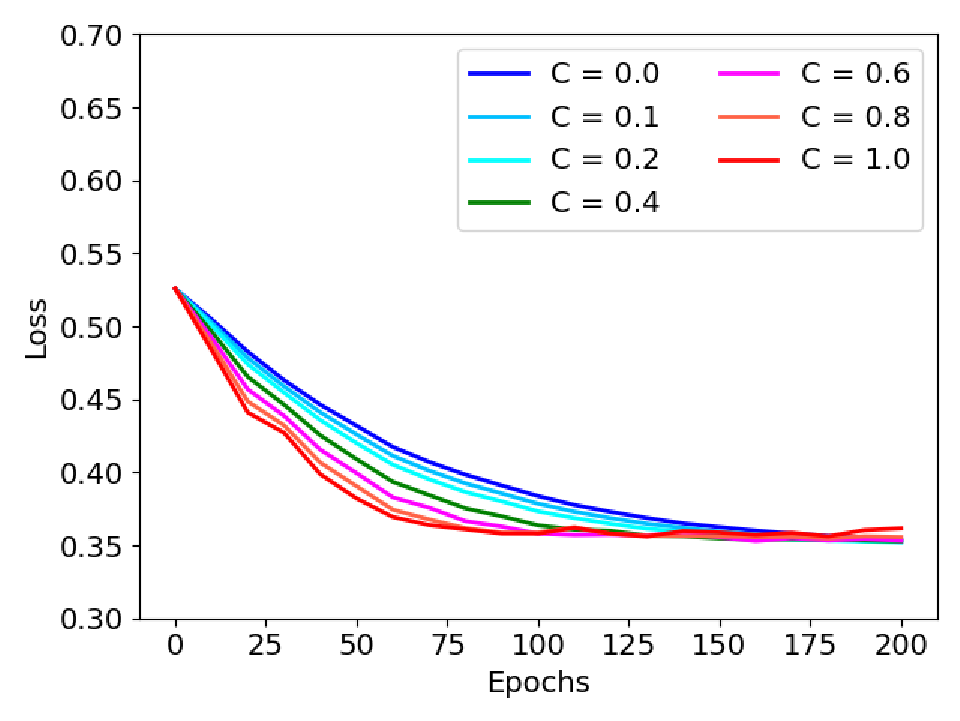}}
    \subfigure[d = 2 (t), up to 200 epochs]{\includegraphics[width=5cm]{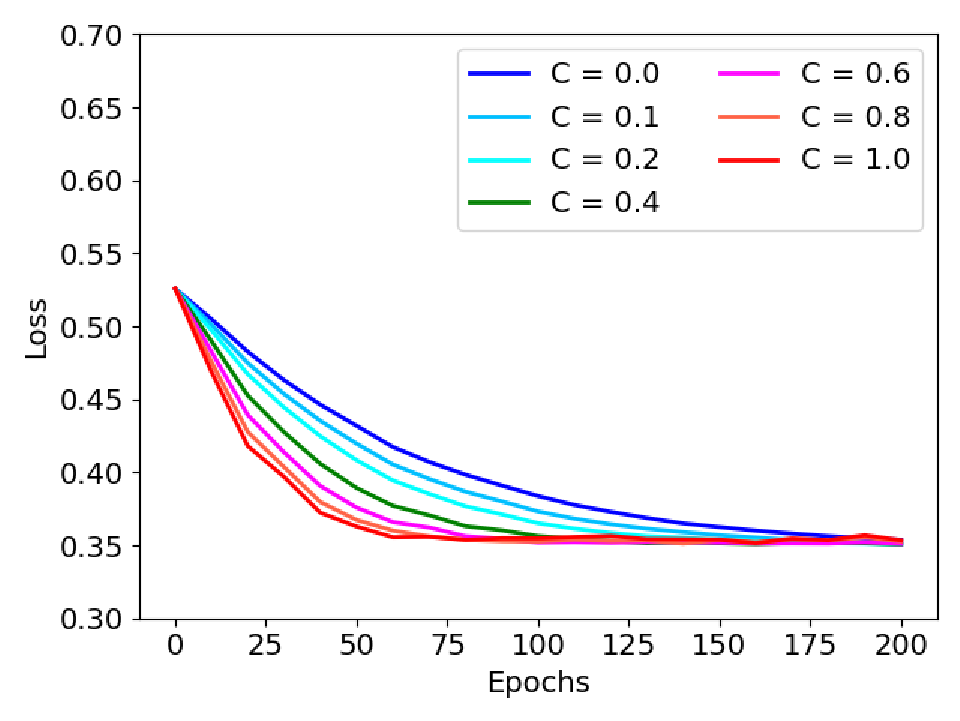}}\\
    \subfigure[d = 4 (e), up to 200 epochs]{\includegraphics[width=5cm]{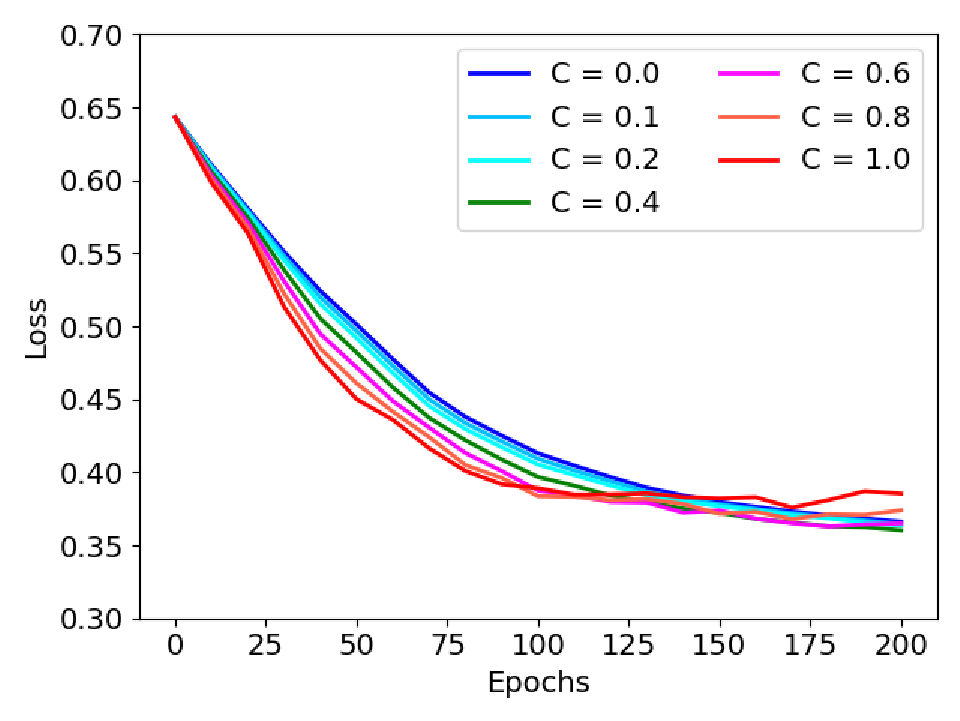}}
    \subfigure[d = 4 (t), up to 200 epochs]{\includegraphics[width=5cm]{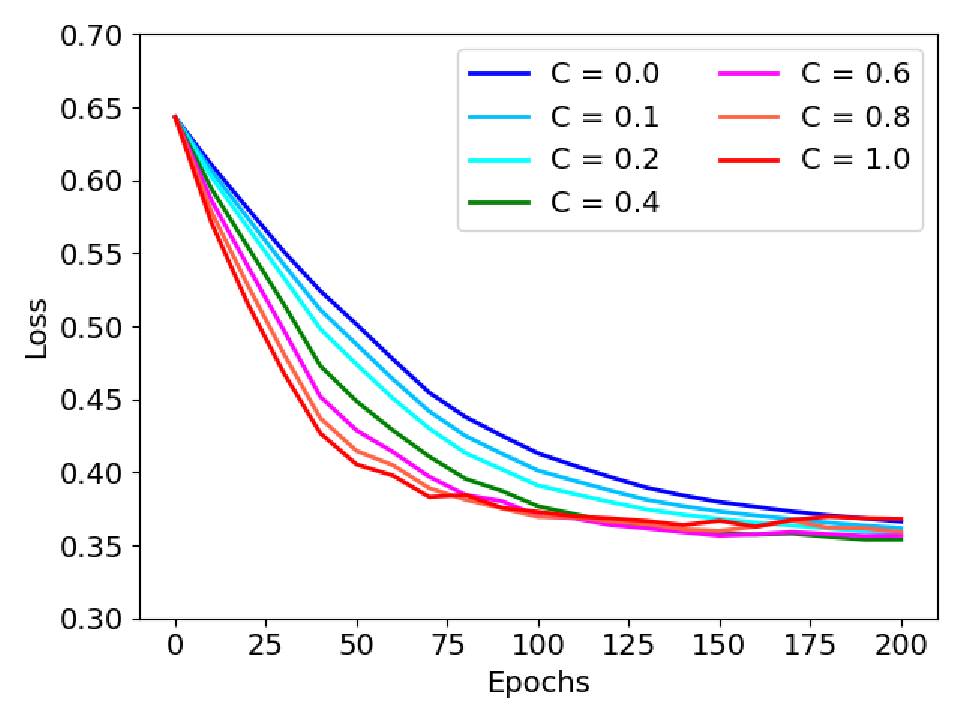}}
  \end{minipage}
  \caption{Loss results for classification tasks (depth 2 and 4). (e) denotes the results using the empirical $ \rho_{in} $ ($ M = 100 $): (t) denotes the results using the theoretical $ \rho_{in} $ (Eq. \ref{eq11}). $ C $ denotes a balancing parameter.}\label{fig4-1}
\end{figure}

The loss results ($ C_{f} $) for depths 2 and 4 are shown in Figure \ref{fig4-1}.
The simulation experiments were performed 20 times with different random seeds.
The results up to 1000 epochs are shown in Figure \ref{fig-a-1} of Appendix \ref{ap3}.
A portion of the results up to 200 epochs is also presented to highlight the behavior in the early training phase.
The results demonstrate that increasing the value of $ C $ enhances the training speed.
While large $ C $ values improved training speed, they resulted in slightly higher converged loss values compared to small $ C $ values.
Regarding the classification accuracy, since the training losses for the examined values of $ C $ were nearly identical at 200 epochs, the accuracy results for $ C = 0.2 $ are reported as a representative example.
For $ C = 0.2 $, the mean training and test accuracies at 200 epochs, averaged over 20 independent runs, were 0.861 (0.0030) and 0.875 (0.0077), respectively, for d = 2 (e); 0.860 (0.0032) and 0.882 (0.0051) for d = 2 (t); 0.865 (0.0074) and 0.872 (0.0100) for d = 4 (e); and 0.861 (0.0077) and 0.875 (0.0088) for d = 4 (t) (other results are shown in Appendix \ref{ap3b}).
Here, the values in parentheses denote one standard deviation.
For the results using the theoretical $ \rho_{in} $, the loss decreased more rapidly than with the empirical $ \rho_{in} $ as $ C $ increased.
For large values of $ C $ (0.8 and 1.0) and in the empirical $ \rho_{in} $ cases, the convergence performance deteriorated somewhat due to excessive regularization.
However, using the theoretical $ \rho_{in} $ slightly improved the convergence performance.
Regarding the minimum loss at 1000 epochs, there were significant differences (p-value $ < $ 0.01) between the results for d = 2 and d = 4 ($ C \in \{0.4, 0.6, 0.8, 1.0\} $), regardless of whether the empirical or theoretical $ \rho_{in} $ was used\footnote{
  Since the number of data points was 20 for statistical testing, we conducted the Shapiro-Wilk normality test on the data.
  Depending on the results, either a two-sided Student's t-test was used (when normality was not rejected) or the Wilcoxon signed-rank test (when normality was rejected).
}.
The loss values for d = 4 were significantly larger than those for d = 2.

\begin{figure}[htbp]
  \centering
  \begin{minipage}{12cm}
    \centering
    \SetFigLayout{2}{2}
    \subfigure[d = 2 (e)]{\includegraphics[width=5cm]{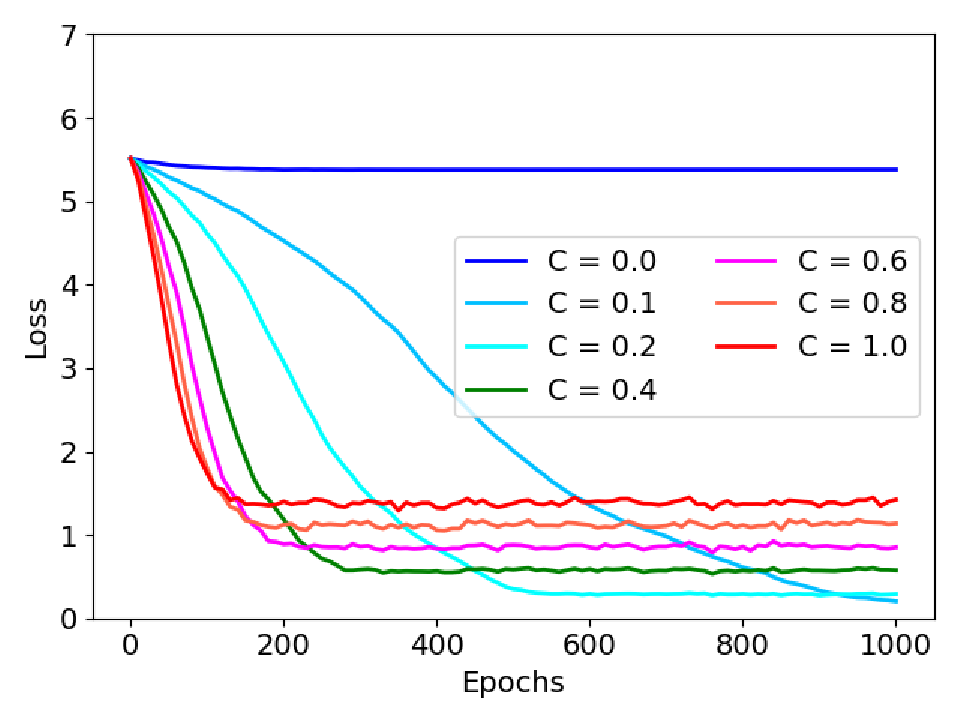}}
    \subfigure[d = 2 (t)]{\includegraphics[width=5cm]{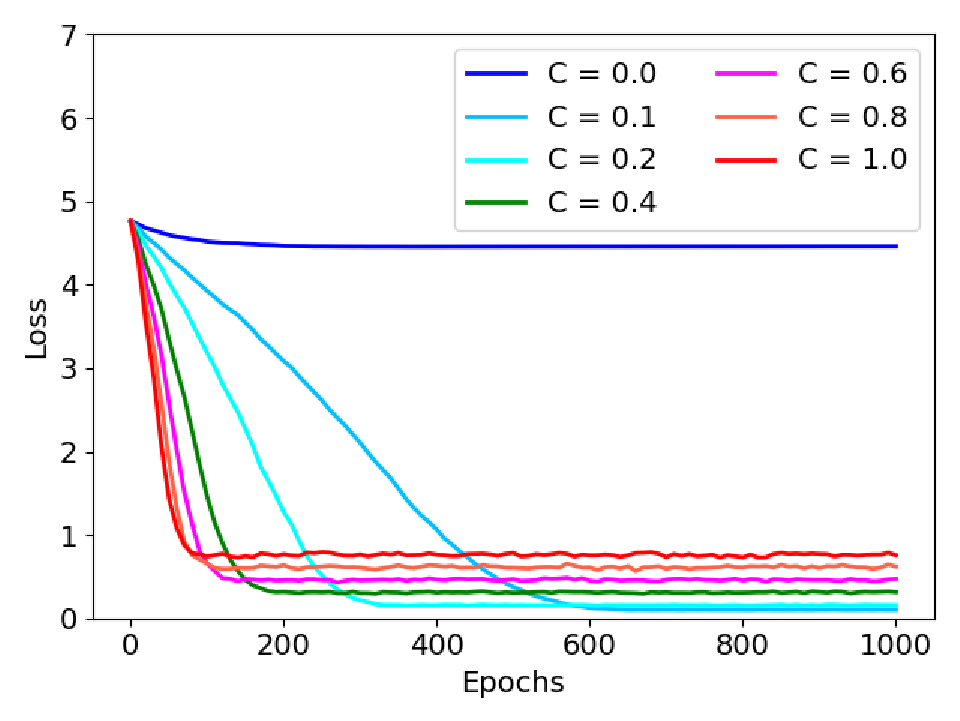}}\\
    \subfigure[d = 4 (e)]{\includegraphics[width=5cm]{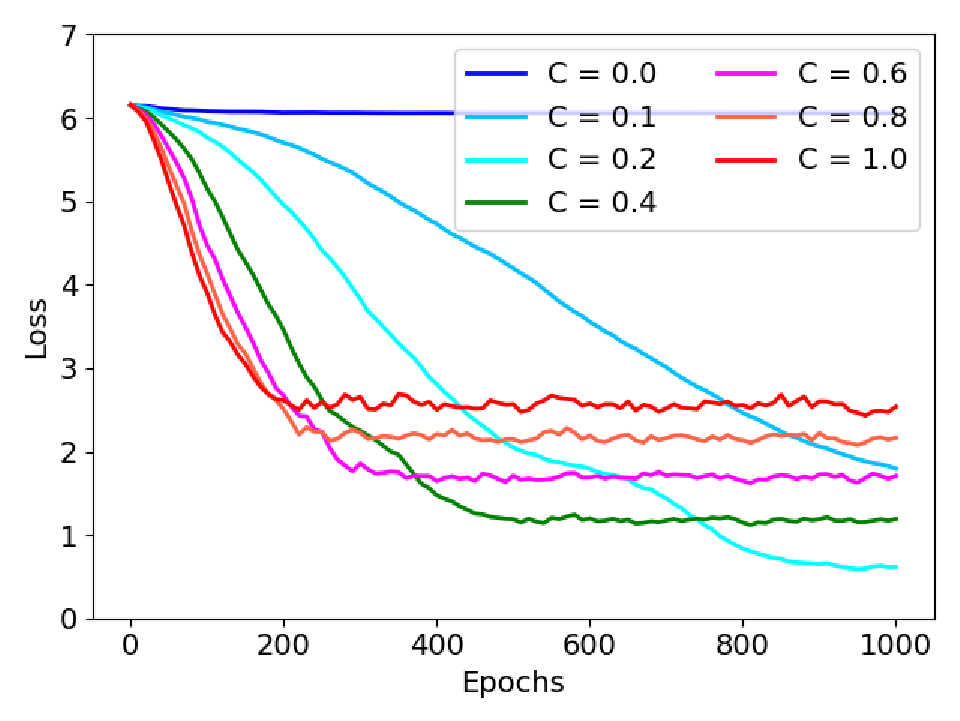}}
    \subfigure[d = 4 (t)]{\includegraphics[width=5cm]{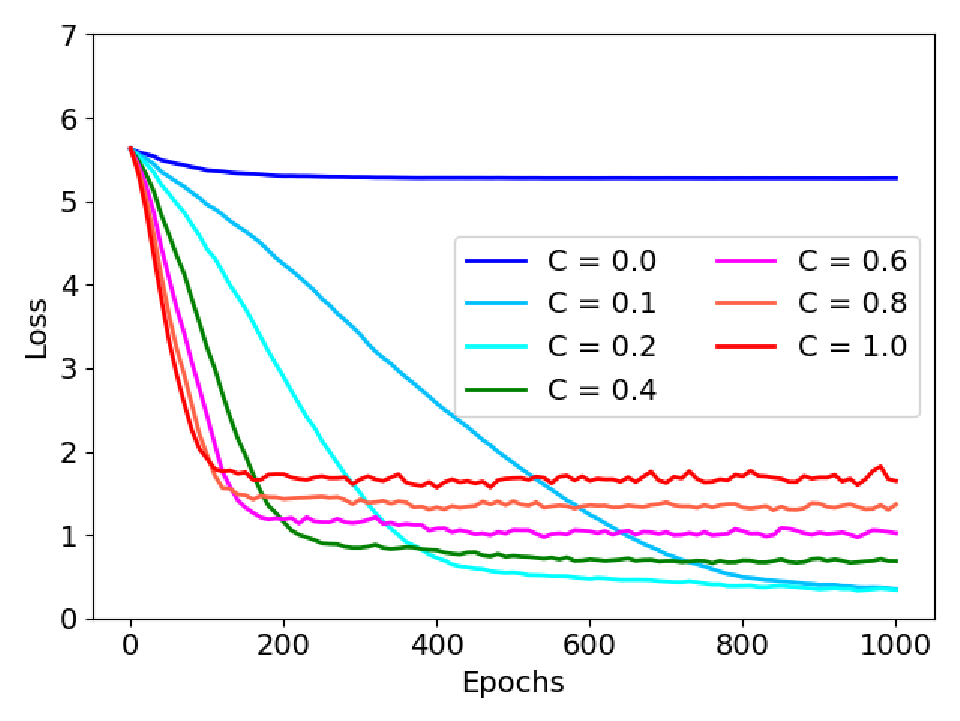}}
  \end{minipage}
  \caption{Loss results for the penalty loss $ C_{symm} $ for classification tasks. (e) denotes the results using the empirical $ \rho_{in} $ ($ M = 100 $); (t) denotes the results using the theoretical $ \rho_{in} $ (Eq. \ref{eq11}).}\label{fig4-2}
\end{figure}

The loss results for the penalty loss term $ C_{symm} $ (Equation \ref{eq19}) are shown in Figure \ref{fig4-2}.
According to the figures, the case with $ C = 0.2 $ achieved the lowest loss among the values of $ C $.
Larger values of $ C $ led to a faster decrease in the loss, but the loss did not decrease sufficiently overall.
Such insufficient loss reduction may be due to local minima.
The results using the theoretical $ \rho_{in} $ exhibited slightly lower loss values.
Additionally, it was observed that even when $ C = 0 $, the penalty loss decreased slightly.

\begin{figure}[htbp]
  \centering
  \begin{minipage}{12cm}
    \centering
    \SetFigLayout{2}{2}
    \subfigure[d = 2 (e)]{\includegraphics[width=5cm]{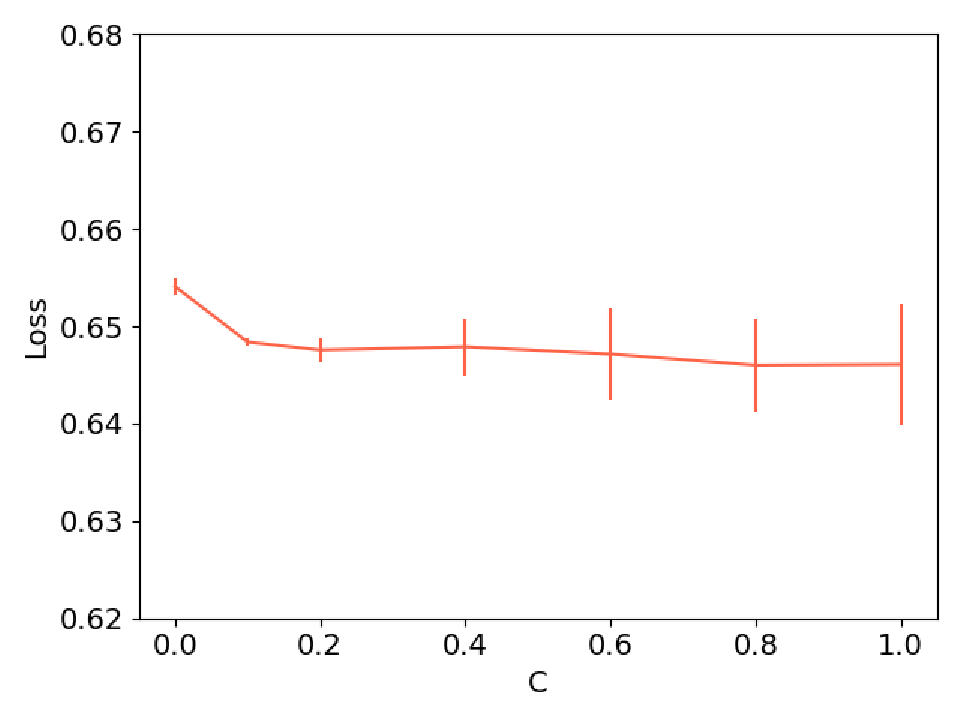}}
    \subfigure[d = 2 (t)]{\includegraphics[width=5cm]{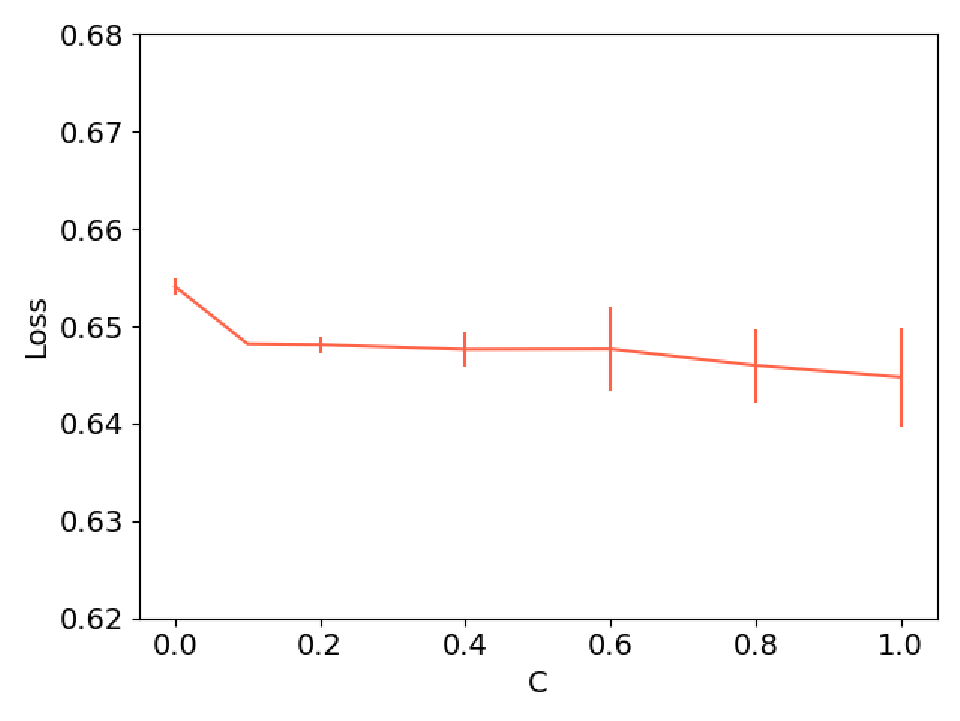}}\\
    \subfigure[d = 4 (e)]{\includegraphics[width=5cm]{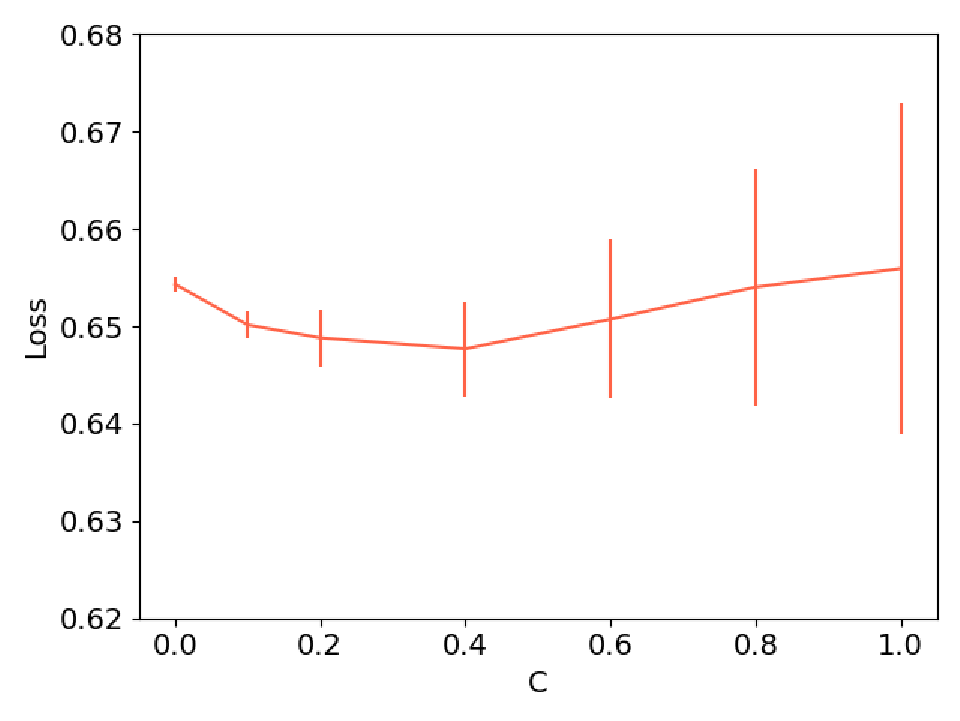}}
    \subfigure[d = 4 (t)]{\includegraphics[width=5cm]{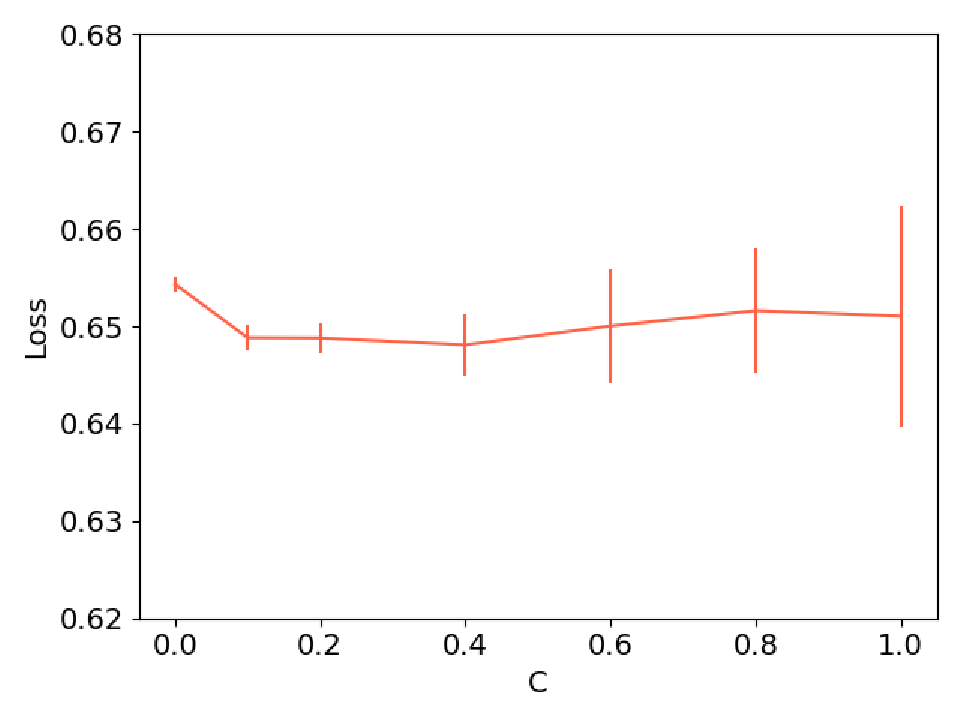}}
  \end{minipage}
  \caption{Test loss results for classification tasks with respect to a balancing parameter $ C $ ($ \in \{0, 0.1, 0.2, 0.4, 0.6, 0.8, 1.0\} $). (e) denotes the results using the empirical $ \rho_{in} $ ($ M = 100 $). (t) denotes the results using the theoretical $ \rho_{in} $ (Eq. \ref{eq11}). Error bar denotes a standard deviation.}\label{fig4-3}
\end{figure}

Next, the test loss results (i.e., generalization performance) are shown in Figure \ref{fig4-3}.
The results indicate that increasing the value of $ C $ leads to a decrease in test loss, thereby improving generalization performance.
Furthermore, the variance of the test loss appeared to increase with larger $ C $ values, likely due to convergence to local minima.
Among all cases, the combination of d = 2 (t) and $ C = 1.0 $ yielded the lowest test loss.
Thus, there were significant differences (p-value $ < $ 0.01) between the test losses $ C = 0 $ and $ C = 1.0 $.

\subsection{Comparison to weight decay method}
\begin{figure}[htbp]
  \centering
  \begin{tabular}{c}
    \includegraphics[width=5cm]{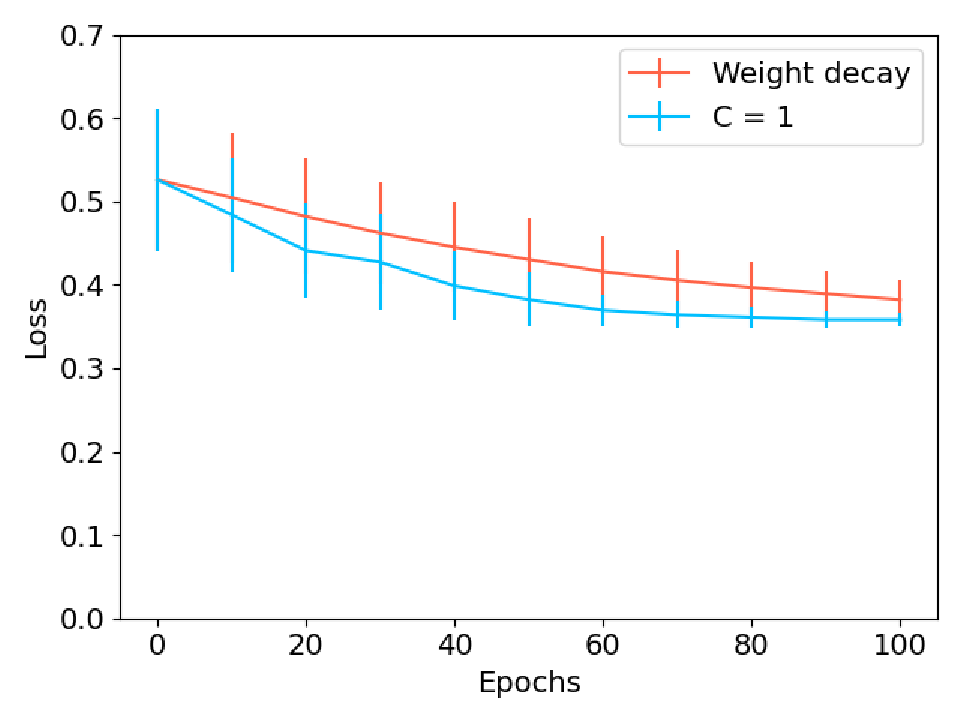}
  \end{tabular}
  \caption{Comparison results of the weight decay method and the proposed method ($ C = 1 $). Error bar denotes a standard deviation.}\label{fig4-3-1}
\end{figure}

We compared the proposed method ($ C = 1 $) with the weight decay method, which is a widely used regularization technique that does not require prior symmetry knowledge.
Here, the weight decay rate was $ 10^{-4} $.\footnote{
  We tuned the weight decay rate over \{1e-2, 1e-3, 1e-4, 1e-5, 1e-6\} and report the value that achieved the best performance in terms of training and test losses.
}
The comparison results of the average loss ($ C_{f} $) over 20 times are shown in Figure \ref{fig4-3-1}.
The error bars indicate one standard deviation.
As shown in Figure \ref{fig4-3-1}, the loss decrease was improved in the early stage of training, comparing to the weight decay method.
However, as increasing epochs, the loss differences gradually became small.
At 100 epochs, the average test loss over 20 times was 0.6444 (0.0092) for the proposed method and 0.6730 (0.0260) for the weight decay method where the standard deviations are shown in parentheses.
There was a significant difference between them (p-value $ < $ 0.01).
Therefore, this implies that the regularization effect of the proposed method is at least comparable to, and possibly slightly better than, that of the weight decay method.

\subsection{Generative task results}
\begin{figure}[htbp]
  \centering
  \begin{minipage}{12cm}
    \centering
    \SetFigLayout{2}{2}
    \subfigure[Negative loss]{\includegraphics[width=5cm]{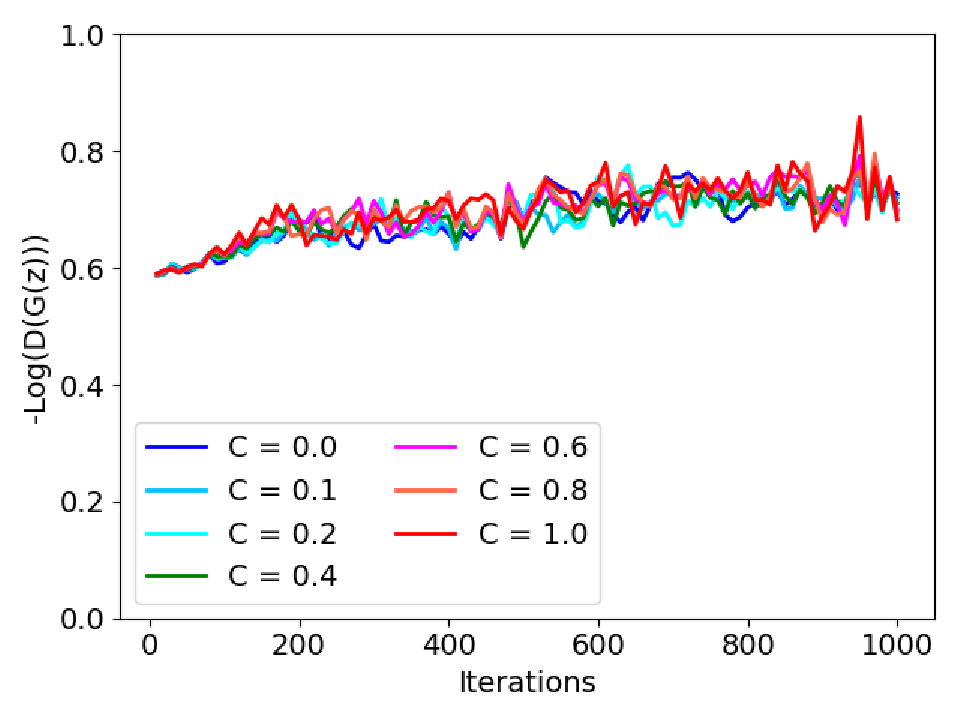}}
    \subfigure[Negative loss, up to 200 iterations]{\includegraphics[width=5cm]{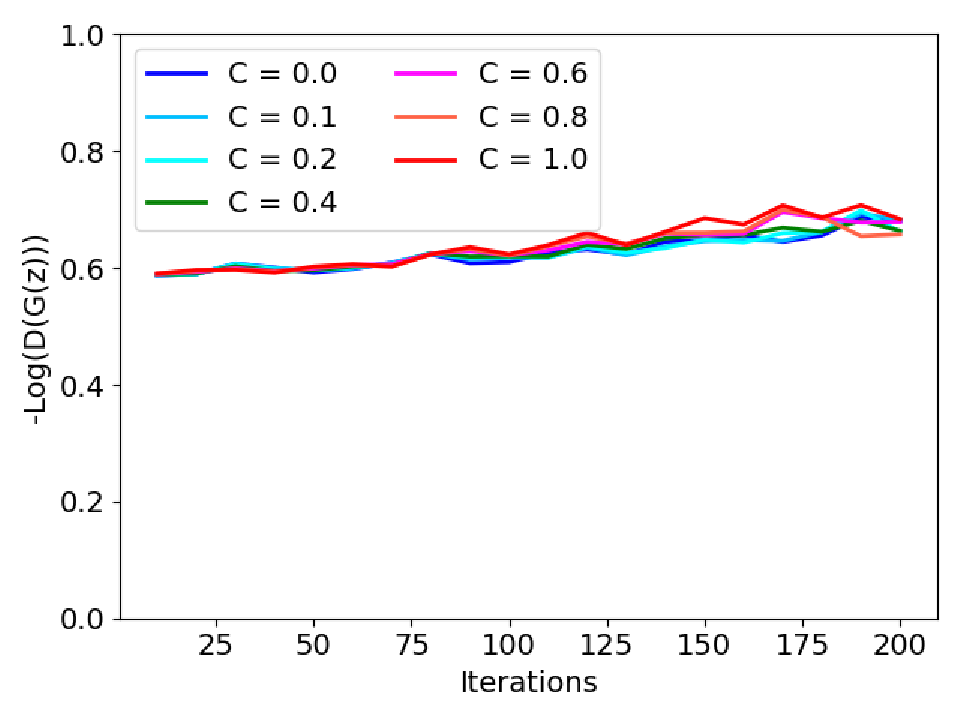}}\\
    \centering
    \subfigure[Penalty loss]{\includegraphics[width=5cm]{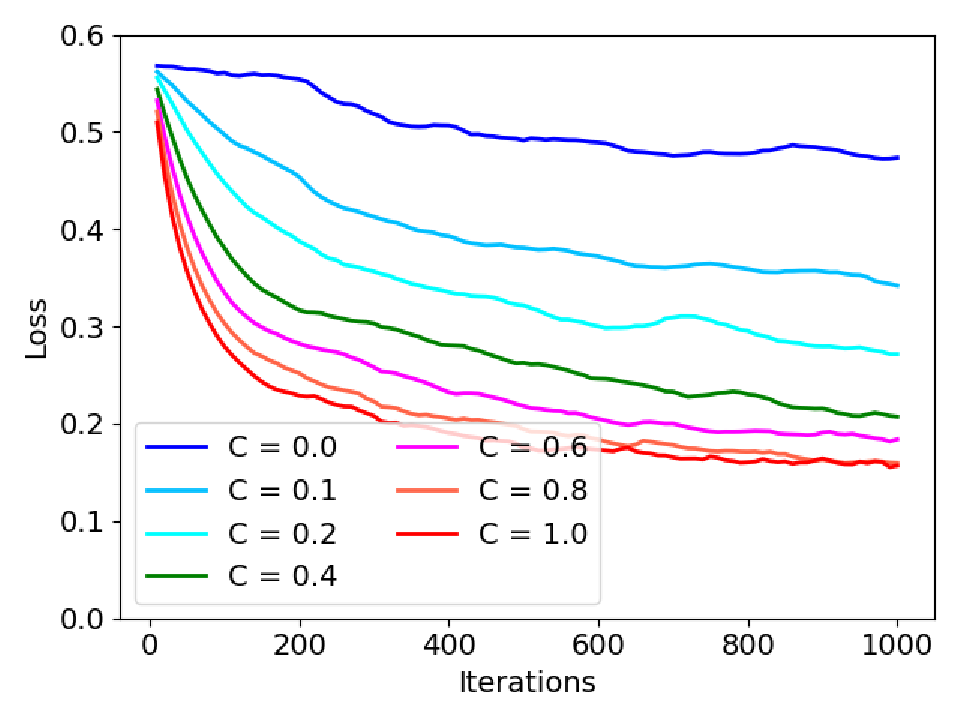}}
  \end{minipage}
  \caption{Loss results of the generator (G) in QGAN for generative tasks. The theoretical $ \rho_{in} $ (Eq. \ref{eq10}) is used.}\label{fig4-4}
\end{figure}

Figure \ref{fig4-4} shows the negative loss ($ -\log(D(G(z))) $) of G in QGAN and the loss of the penalty term. 
From the figure, it can be observed that increasing $ C $ led to a slight increase in the negative loss, while the penalty loss clearly decreased as $ C $ increased.
Moreover, Fr\'{e}het inception distance (FID) \cite{szegedy2016} was used for the evaluation of the generated images.
At 500 epochs, the average FID value over 10 times was 353.95 (21.91) for $ C = 0 $ and 354.66 (20.87) for $ C = 1 $, where the standard deviations are shown in parentheses.
There was no significant difference between them (p-value = 0.942).
However, as shown in Figure \ref{fig-a-3} of Appendix \ref{ap4}, in the early stage of training up to 100 iterations, the mean FID curve for $ C = 1 $ was lower during part of the early training stage, but the difference was not established as statistically significant.

\begin{figure}[htbp]
  \centering
  \begin{minipage}{12cm}
    \centering
    \SetFigLayout{1}{2}
    \subfigure[C = 0]{\includegraphics[width=5cm]{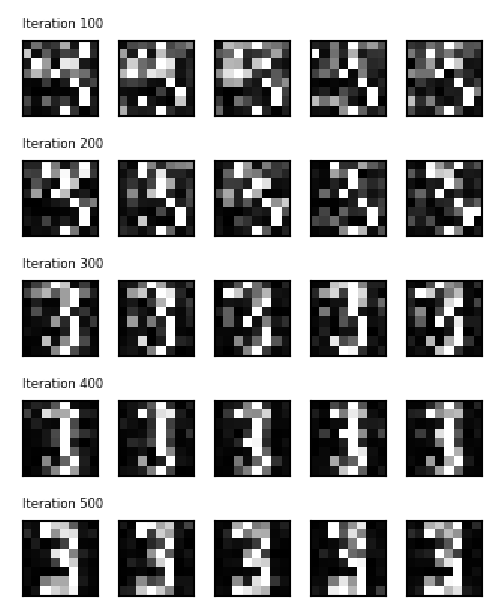}}
    \hfill
    \subfigure[C = 1]{\includegraphics[width=5cm]{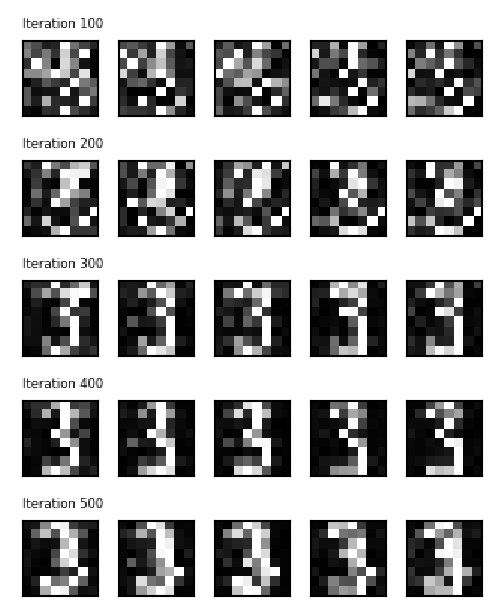}}
  \end{minipage}
  \caption{Sampling results of quantum generators for $ C = 0 $ (without data symmetry) and $ C = 1 $ (with data symmetry) during training. Five samples are shown at every 100 iterations.}\label{fig4-5}
\end{figure}

Figure \ref{fig4-5} presents the sampling results every 100 iterations during training.
Five samples were generated by the generator of QGAN.
Comparing the results for $ C = 0 $ and $ C = 1 $, the samples at 400 iterations with $ C = 1 $ appeared slightly clearer than those with $ C = 0 $.
The QGAN results provide only preliminary qualitative evidence of an early-stage effect; no statistically significant improvement was observed in FID.

\section{Conclusion}\label{sec4}
In this study, we proposed a heuristic symmetry-informed regularization method based on the input density matrix.
The method constrains the model search space toward unitaries commuting with a symmetry operator selected from the commutant of the input density matrix.
The proposed method introduces a penalty term related to data symmetry into the cost function during training.
A balancing parameter for the penalty term controls the regularization effects during training.
We conducted two types of experiments: classification and generative tasks.
The classification experiments showed modest improvements in early-stage convergence and test loss.
In contrast, the QGAN experiment did not show a statistically significant improvement in the final FID, indicating that the benefit of the regularizer is task- and ansatz-dependent.
The proposed method is simple and easy to implement.
Nonetheless, several issues remain.
For large qubit sizes, the spectral decomposition used in the $ S $ calculation algorithm may become computationally challenging.
Additionally, the choice of a classical data encoding method is an important issue for the proposed method, particularly with respect to the construction of $ \rho_{in} $ and the training process (i.e., optimization speed).
Future work should address these issues.

\appendix
\section{An algorithm for calculating $ S $}\label{ap1}
Given $ U $ (unitary or Hermitian matrix of size $ N $), the following algorithm gives $ S $ such that $ S U S^{\dagger} = U $.
\begin{enumerate}
\item Perform the spectral decomposition of $ U $:\\ $ U = V D V^{\dagger} $, where $ D = {\rm diag}(e^{i \lambda_{1}}, e^{i \lambda_{2}}, \cdots, e^{i \lambda_{N}}) $ ($ \lambda_{j} $ is the eigenvalue of $ U $ in the case of non-degenerate eigenvalues).
\item Construct $ S $:\\ $ S = V \, {\rm diag}(e^{i \phi_{1}}, e^{i \phi_{2}}, \cdots, e^{i \phi_{N}}) \, V^{\dagger} $, where $ \phi_{j} $ denotes an arbitrary real number.
\end{enumerate}
In the case of degenerate eigenvalues, any unitary matrix of size $ k $ corresponding to the degenerate eigenvalues can be placed as a diagonal block in $ D $, where $ k $ denotes the degree of degeneracy.
The computational complexity of this algorithm is $ \mathcal{O} (N^{3}) $.
In this study, for $ \phi $s, random values followed by the standard normal distribution were used.

\section{Mismatch bound}\label{ap2}
When the estimated $ S $ differs from the true symmetry matrix $ S_{true} $, e.g., $ S = S_{true} + \delta S $, where $ \delta S $ is a mismatch matrix (which is not necessarily unitary).
$ S W S^{\dagger} - W = (S_{true} W S_{true}^{\dagger} - W) + \delta S W S_{true}^{\dagger} + S_{true} W \delta S^{\dagger} + \delta S W \delta S^{\dagger} $.
Here, we use $ \| A B \}_{{\rm F}} \leq \| A \|_{{\rm F}} \| B \|_{2} $.
Hence, since $ W $ and $ S_{true} $ are unitary, an upper-bound on the regularization term is given by
\begin{equation*}
  \| S W S^{\dagger} - W \|_{{\rm F}} \leq \| S_{true} W S_{true}^{\dagger} - W \|_{{\rm F}} + 2 \| \delta S \|_{{\rm F}} + \| \delta S \|^{2}_{{\rm F}}.
\end{equation*}
This bound quantifies the possible increase in the symmetry-regularization residual caused by a mismatch in $ S $.
It does not, by itself, imply degradation of the task loss or generalization performance.

\section{Classification task results up to 1000 epochs}\label{ap3}
\begin{figure}[htbp]
  \centering
  \begin{minipage}{8.8cm}
    \centering
    \SetFigLayout{2}{2}
    \subfigure[d = 2 (e)]{\includegraphics[width=3.5cm]{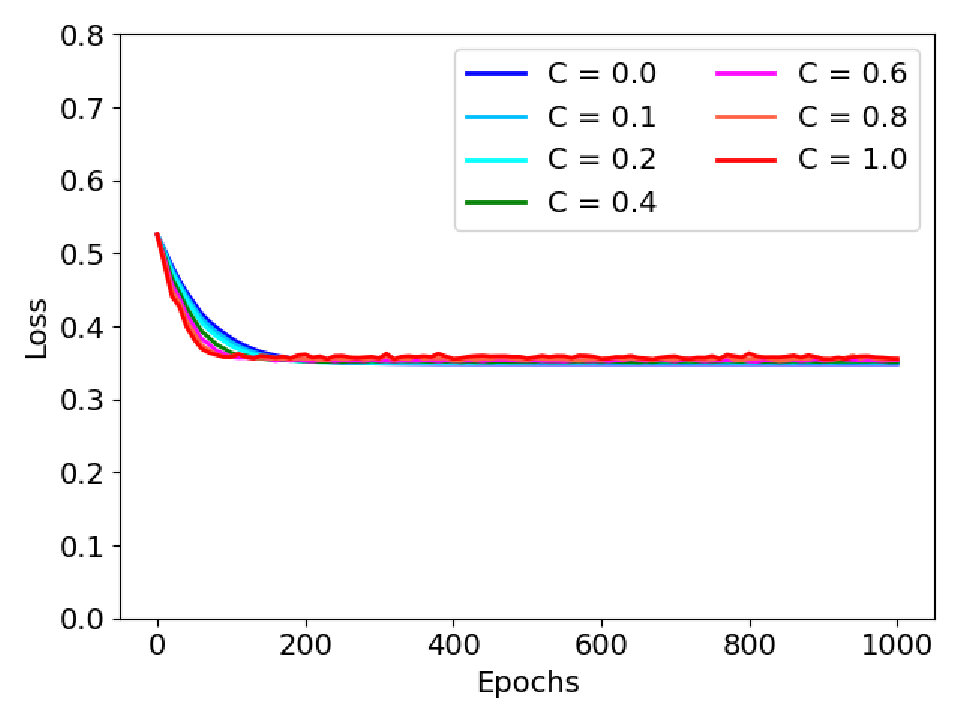}}
    \subfigure[d = 2 (t)]{\includegraphics[width=3.5cm]{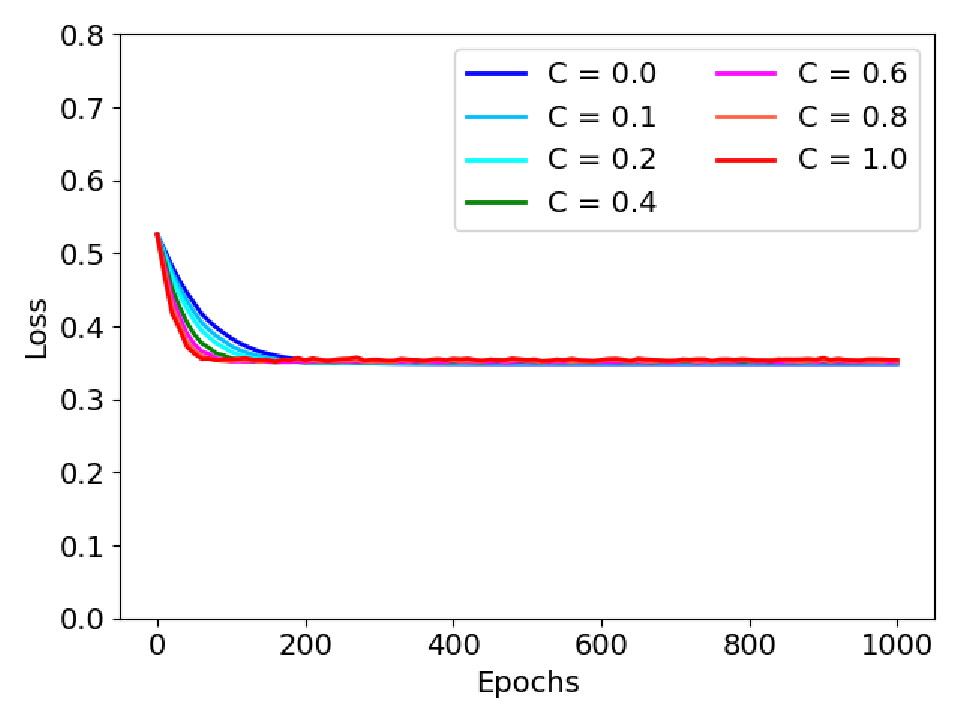}}\\
    \subfigure[d = 4 (e)]{\includegraphics[width=3.5cm]{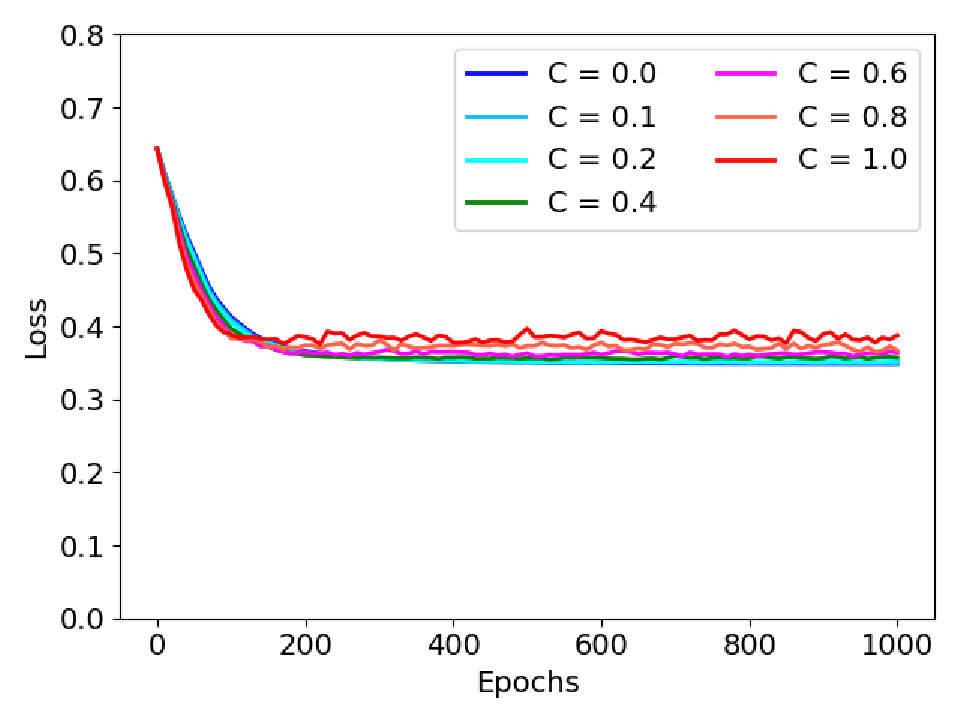}}
    \subfigure[d = 4 (t)]{\includegraphics[width=3.5cm]{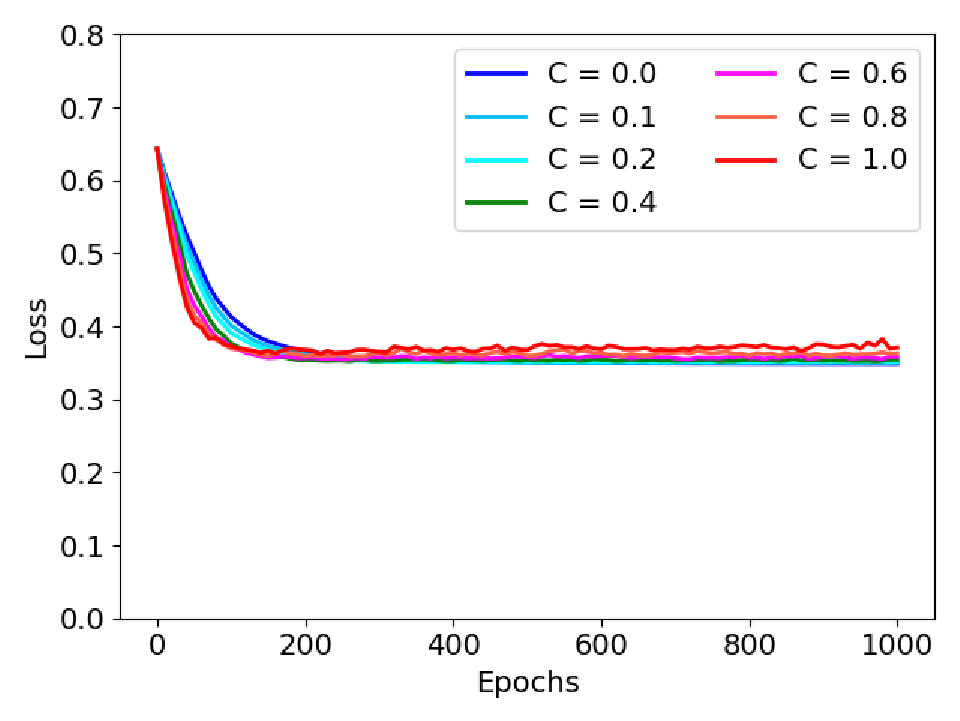}}
  \end{minipage}
  \caption{Loss results for classification tasks (depth 2 and 4). (e) denotes the results using the empirical $ \rho_{in} $ ($ M = 100 $): (t) denotes the results using the theoretical $ \rho_{in} $ (Eq. \ref{eq11}). $ C $ denotes a balancing parameter.}\label{fig-a-1}
\end{figure}

Figure \ref{fig-a-1} shows the loss results up to 1000 epochs.

\section{Classification accuracy results}\label{ap3b}
\begin{table}[H]
  \centering
  \caption{Results of classification accuracy}\label{tabap1}
  \vspace{5pt}
  \begin{tabular}{ccc|ccc}\hline
    $ C = 0.0 $ & Training accuracy & Test accuracy & $ C = 0.2 $ & Training accuracy & Test accuracy\\ \hline
    d = 2 (e) & 0.867 (0.0079) & 0.869 (0.0054) & d = 2 (e) & 0.861 (0.0030) & 0.875 (0.0077)\\
    d = 2 (t) & 0.867 (0.0079) & 0.869 (0.0054) & d = 2 (t) & 0.860 (0.0032) & 0.882 (0.0051)\\
    d = 4 (e) & 0.865 (0.0074) & 0.869 (0.0059) & d = 4 (e) & 0.865 (0.0074) & 0.872 (0.0100)\\
    d = 4 (t) & 0.865 (0.0074) & 0.869 (0.0059) & d = 4 (t) & 0.861 (0.0077) & 0.875 (0.0088)\\ \hline
    $ C = 0.4 $ & Training accuracy & Test accuracy & $ C = 0.6 $ & Training accuracy & Test accuracy\\ \hline
    d = 2 (e) & 0.860 (0.0050) & 0.878 (0.0075) & d = 2 (e) & 0.864 (0.0080) & 0.881 (0.0109)\\
    d = 2 (t) & 0.861 (0.0054) & 0.883 (0.0066) & d = 2 (t) & 0.863 (0.0062) & 0.881 (0.0074)\\
    d = 4 (e) & 0.857 (0.0091) & 0.873 (0.0060) & d = 4 (e) & 0.861 (0.0097) & 0.872 (0.0085)\\
    d = 4 (t) & 0.861 (0.0083) & 0.879 (0.0069) & d = 4 (t) & 0.865 (0.0086) & 0.874 (0.0111)\\ \hline
    $ C = 0.8 $ & Training accuracy & Test accuracy & $ C = 1.0 $ & Training accuracy & Test accuracy\\ \hline
    d = 2 (e) & 0.863 (0.0070) & 0.879 (0.0117) & d = 2 (e) & 0.861 (0.0080) & 0.878 (0.0081)\\
    d = 2 (t) & 0.862 (0.0065) & 0.877 (0.0097) & d = 2 (t) & 0.861 (0.0083) & 0.880 (0.0100)\\
    d = 4 (e) & 0.858 (0.0093) & 0.869 (0.0110) & d = 4 (e) & 0.857 (0.0105) & 0.872 (0.0104)\\ \hline
  \end{tabular}
\end{table}

\section{Generative task results up to 100 iterations}\label{ap4}
\begin{figure}[htbp]
  \centering
  \begin{tabular}{c}
    \includegraphics[width=5cm]{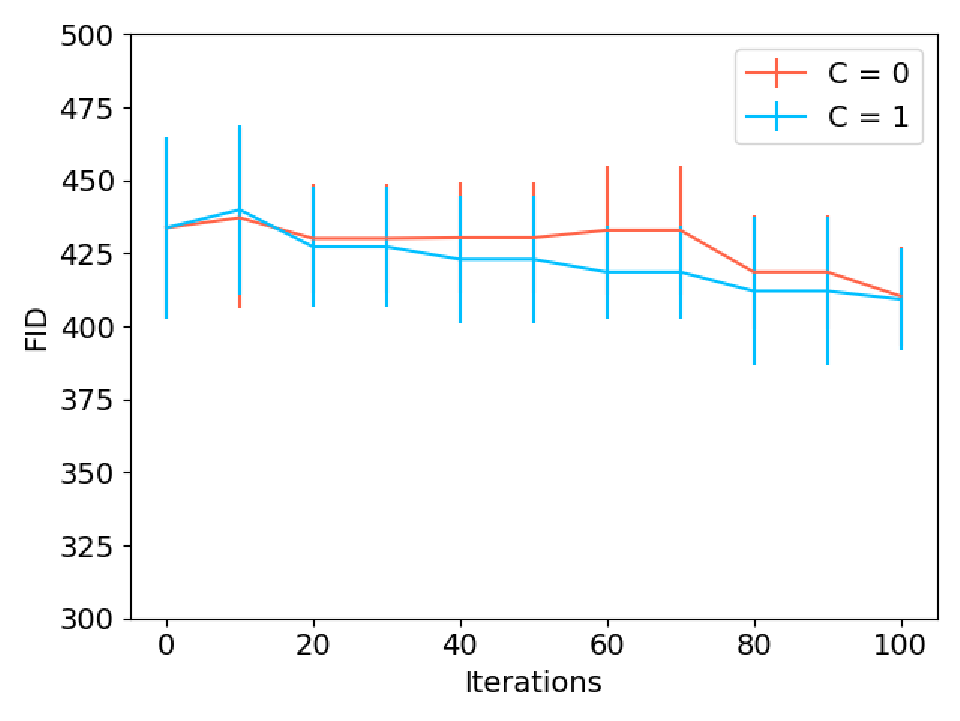}
  \end{tabular}
  \caption{FID results up to 100 iterations}\label{fig-a-3}
\end{figure}

FID results up to 100 iterations are shown in Figure \ref{fig-a-3}.
The error bars indicate one standard deviation.

\bibliographystyle{spmpsci}      
\bibliography{references-r2}   

\end{document}